\newcommand{\ket}[1]{|#1\rangle}

\documentclass[conference]{IEEEtran}
\usepackage{graphicx}
\usepackage{amsmath}
\usepackage{epstopdf}
\usepackage{comment}

\usepackage{cite}
\usepackage{amsmath,amssymb,amsfonts}
\usepackage{algorithmic}
\usepackage{graphicx}
\usepackage{textcomp}

\usepackage{amssymb}
\usepackage{graphicx}
\usepackage{amsmath}
\usepackage{algorithmic}
\usepackage{graphicx}
\usepackage{graphics}
\usepackage{url}
\usepackage{blkarray}
\usepackage{mathtools}
\usepackage{float}
\usepackage{multicol}
\usepackage{cite}
\usepackage{graphicx}
\usepackage{epstopdf}
\usepackage{epsfig}
\usepackage{stmaryrd}
\usepackage{gensymb}
\usepackage[utf8]{inputenc}
\usepackage{color}
\usepackage{multirow}
\usepackage[ruled,vlined]{algorithm2e}

\IncMargin{-\parindent}
\DeclarePairedDelimiter{\ceil}{\lceil}{\rceil}

\begin{document}
%
\title{Circuit Design for $k$-coloring Problem and Its Implementation on Near-term Quantum Devices}

\author{
\IEEEauthorblockN{Amit Saha$^{1,2}$, Debasri Saha$^1$, and Amlan Chakrabarti$^1$}
\IEEEauthorblockA{$^1$A. K. Choudhury School of Information Technology,  University of Calcutta, Kolkata - 700 106, India}
{$^2$ATOS, Pune, India}
}


%


\maketitle

\begin{abstract}
Nowadays in Quantum Computing, the implementation of quantum algorithm has created a stir since Noisy Intermediate-Scale Quantum (NISQ) devices are out in the market. Researchers are mostly interested in solving NP-complete problems with the help of quantum algorithms for its speed-up. As per the work on computational complexity by Karp \cite{karp}, if any of the NP-complete problem can be solved then any other NP-complete problem can be reduced to that problem in polynomial time. In this Paper, $k$-coloring problem (NP-complete problem) has been considered to solve using Grover's search. A comparator-based approach has been used to implement $k$-coloring problem which enables the reduction of the qubit cost compared to the state-of-the-art. An end-to-end automated framework has been proposed to implement the $k$-coloring problem for any unweighted and undirected graph on any available Noisy Intermediate-Scale Quantum (NISQ) devices, which helps in generalizing our approach.  
\end{abstract}



%
\IEEEpeerreviewmaketitle

\section{Introduction}

As the development of Noisy Intermediate-Scale Quantum (NISQ) computer \cite{preskil} has achieved a remarkable success in recent times, everyone has shown a striking interest to implement quantum algorithms, which give a potential speedup over their classical counterparts. With the growing quantum wave, there is a huge urge for implementing NP-complete problems on near term quantum devices. It would be helpful for any naive person, if we could provide them with an automated end-to-end framework for implementing an NP-complete problem so that they can easily map their computational problem without having much knowledge about  gate-based quantum circuit implementation. In this paper, we have focused on $k$-coloring problem.

\par  The $k$-coloring problem finds whether a given graph's vertices or nodes are properly colored or not using $k$ colors by taking into account that every two vertices linked by an edge have different colors. Suppose $n$ is the number of nodes of a given graph, $k$ is the number of colors, then to find the exact solution using a classical algorithm requires $O(2^{n*log k})$ number of steps. Whereas, using the decision oracle and the diffusion operator of  Grover's algorithm \cite{Grover's}, finding the exact solution requires $O\sqrt{N}$ number of iterations where $N$ is $2^{n*log k}$. Previously in \cite{amit} \cite{graph_color}, Graph coloring problem using Grover's algorithm has been discussed in the context of quantum system. But, in \cite{ibm_graphcolor} SAT reduction technique has been used to solve 3-coloring problem and gave an end-to-end framework for implementing it in the IBMQ quantum processor \cite{4}. For this SAT reduction technique, the qubit cost is immense, hence circuit cost becomes inefficient.

In this paper, we have proposed an automated qubit cost-efficient comparator-based approach to implement $k$-coloring problem for mapping high level description to any hardware-specific low-level quantum operations with an abstraction. The novelty of this paper is as follows:

\begin{itemize}
    \item We propose an end-to-end automated framework for $k$-coloring problem using quantum search algorithm, which takes graph and number of color ($k$) as input and automatically implements on the NISQ device.
    \item We propose a comparator-based approach to implement the $k$-coloring problem which has less qubit cost comparing to the state-of-the-art.
    \item The framework is designed in such a way that the Quantum solution of $k$-coloring problem can be mapped into any available NISQ devices, which makes our approach generalized in nature.
\end{itemize}

\par The structure of this paper is as follows. The synopsis of Grover's algorithm, Quantum circuits, and NISQ devices are described in section II. In section III, the proposed methodology has been discussed. The implementation of $k$-coloring problem has been illustrated in section IV. Concluding remarks appear in Section V.

\section{Background}

In this section, we have mainly described about quantum circuit, Grover's algorithm and finally NISQ devices.

\subsection{Quantum circuit}
Any quantum algorithm can be expressed or visualized in the form of a quantum circuit. These quantum circuits constitute of logical qubits and quantum gates \cite{shor02}.
\subsubsection{Qubits}
 Logical qubit that is used to encode
input or output of a quantum algorithm is known as data qubit. There is an another type of qubit that is used to store temporary results are known as ancilla qubit.
\subsubsection{Quantum Gates}
 Unitary quantum gates need to be applied on qubits to modify the quantum state of a quantum algorithm. To synthesize our proposed circuit, we use NOT gate, Controlled-NOT gate, Toffoli gate, Hadamard gate and Multi Control Toffoli gate(MCT). All the mentioned gates except the MCT are described in Table \ref{table1}. The description of MCT gate is as follows:

\begin{table}[!h]
\caption{Matrix and Circuit Representation of Quantum Gates}
\centering
\resizebox{7cm}{!}{%
\begin{tabular}{ | c | c | c |}

 \hline
 Quantum Gates & Matrix Representation & Circuit Representation\\
 \hline
 Hadamard Gate & $\begin{pmatrix}
\\ \frac{1}{\sqrt{2}} & \frac{1}{\sqrt{2}}
\\ \frac{1}{\sqrt{2}} & -\frac{1}{\sqrt{2}}
\end{pmatrix}$ & 
\includegraphics[width=30mm]{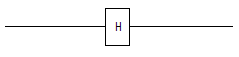}\\
 \hline
 Not Gate  &  $\begin{pmatrix}
\\ 0 & 1
\\ 1 & 0
\end{pmatrix}$ & 
\includegraphics[width=30mm]{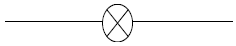}\\
 \hline
 Controlled-NOT Gate & $\begin{pmatrix}
\\ 1 & 0 & 0 & 0
\\ 0 & 1 & 0 & 0
\\ 0 & 0 & 0 & 1
\\ 0 & 0 & 1 & 0
\end{pmatrix}$ & 
\includegraphics[width=30mm, height=1.2cm]{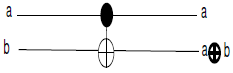}\\
 \hline
 Toffoli Gate &  $\begin{pmatrix}
\\ 1 & 0 & 0 & 0 & 0 & 0 & 0 & 0
\\ 0 & 1 & 0 & 0 & 0 & 0 & 0 & 0
\\ 0 & 0 & 1 & 0 & 0 & 0 & 0 & 0
\\ 0 & 0 & 0 & 1 & 0 & 0 & 0 & 0
\\ 0 & 0 & 0 & 0 & 1 & 0 & 0 & 0
\\ 0 & 0 & 0 & 0 & 0 & 1 & 0 & 0
\\ 0 & 0 & 0 & 0 & 0 & 0 & 0 & 1
\\ 0 & 0 & 0 & 0 & 0 & 0 & 1 & 0
\end{pmatrix}$ & 
\includegraphics[width=30mm, height=1.5cm]{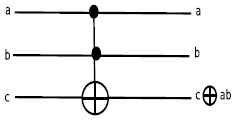}\\
 \hline
\end{tabular}}
\label{table1}
\end{table}

\begin{figure}[!htb]
\centering
\includegraphics[width=60mm, height=1.7cm]{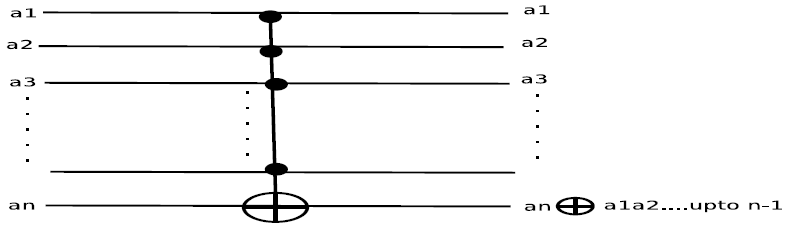}
\caption{Multi-Controlled Toffoli Gate}
\label{mct}
\end{figure}

\textbf{Multi-Controlled Toffoli Gate:} There are n number of inputs and outputs in an $n$-bit MCT. This MCT gate passes the first $n-1$ inputs, which are referred to as control bits to the output unaltered. It inverts the $n^{th}$ input, which is referred to as the target bit if the first $n-1$ inputs are all ones. An MCT gate is shown in Figure \ref{mct} Black dots $\bullet$ represent the control bits and the target bit is denoted by a $\oplus$.


\subsection{Grover's Algorithm}
\begin{figure}[!htb]
\centering
\includegraphics[width=90mm,height=2.0cm]{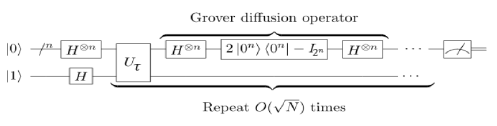}
\caption{Generalized Circuit for Grover's algorithm \cite{Grover's}}
\label{grover}
\end{figure}

Grover's algorithm has two parts, namely oracle and diffusion operator. The oracle depends on the specific instance of the search problem. The diffusion operator block is also known as inversion about the average operator and it amplifies the amplitude of the marked state to increase its measurement probability. The block diagram of a typical Grover's algorithm is shown in Figure \ref{grover}.

\par To perform Grover's search algorithm, at least n + 1 qubits are required and the 
function $f$ is encoded by a unitary $U_{f}$ : $\ket x_{n}$ $\otimes$ $\ket y_{1}$ $\rightarrow$ $\ket x_{n}$ $\otimes$ $\ket {y \oplus f(x)}_{1} $.

\par More elaborately, The steps of the Grover's algorithm are as follows:

\par \textbf{Initialization}: The algorithm starts with the uniform superposition of all the basis states on input qubits $n$. The last ancilla qubit is used as an output qubit which is initialized to $H \ket1$. Thus, we obtain the binary quantum state $\ket \psi$.

\textbf{Sign Flip}: Flip the sign of the vectors for which $U_{f}$ gives output 1.

\textbf{Amplitude Amplification}: We need to perform the inversion about the average of all coefficient of the quantum state for a certain number of iterations to get the coefficient of the marked state is large enough that it can be obtained from a measurement with probability close to 1. This phenomenon is known as amplitude amplification which is performed by using diffusion operator. 

\textbf{Number of Iterations}: Grover's Search algorithm requires $\sqrt{N/M}$ many iterations to get the probability of one of the marked states $M$ out of total $N$ number of states set.

\subsection{NISQ Devices}
\begin{figure}[!h]
\centering
\includegraphics[width=75mm, height=2cm]{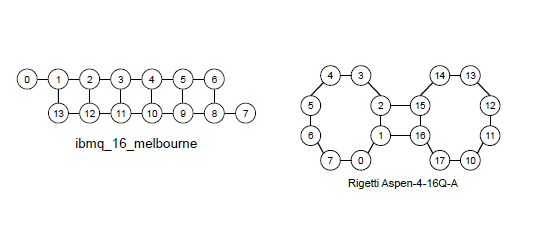}
\caption{Qubit Topology \cite{4, rigetti}}
\label{qt}
\end{figure}

NISQ devices are “noisy,” due to the constraint of the number of qubits, hence one has to allow a certain range of error while estimating the simulated result of a quantum state \cite{preskil}. Superconducting quantum
circuits, ion trap, quantum dot, neutral atom are the most popular NISQ technologies to implement the quantum circuit. Every one of them has a specific qubit topology, as shown in Figure \ref{qt}, so as to map the logical synthesized circuit to quantum hardware. Table \ref{tab:gates} illustrates certain 1-qubit and 2-qubit gates that are supported by most of the quantum hardware. One has to realize their logical quantum gates to these hardware-specific gates to make it hardware compatible for implementation.


\begin{table*}[!h]
\centering
\caption{Gate Set for NISQ Devices}
\begin{tabular}{c|c}
gate type &  gate set \\ \hline
1-qubit gates & id, x, y, z, h, r2, r4, r8, rx, ry, rz, u1, u2, u3, s, t, sdg, tdg\\
2-qubit gates & swap, srswap, iswap, xy, cx, cy, cz, ch, csrn, ms, yy, cr2, cr4, cr8, crx, cry, crz, cu1, cu2, cu3, cs, ct, csdg	
\end{tabular}
\label{tab:gates}
\end{table*}

\section{Proposed Methodology of Circuit Synthesis for $k$-Coloring Problem using Grover's Algorithm}

\begin{figure}[!h]
\centering
\includegraphics[width=45mm, height =5 cm]{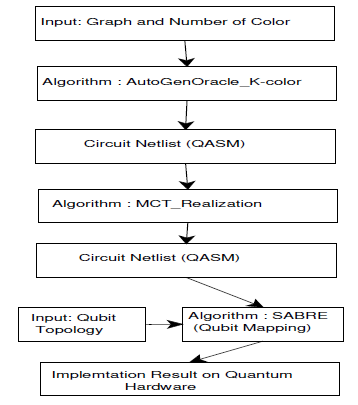}
\caption{Flowchart of our proposed work} 
\label{flow}
\end{figure}

The flowchart as shown in Figure \ref{flow} describes the complete flow of our proposed automated end-to-end framework. Our framework is mainly based on three algorithms: AutoGenOracle\_K-color, MCT\_Realization and SABRE(Qubit Mapping). Firstly, adjacency matrix of the given graph and the number of color ($k$) is given as input to AutoGenOracle\_K-color algorithm and we get the quantum circuit netlist in the form QASM as output. AutoGenOracle\_K-color algorithm automatically generates Oracle circuit for $k$-coloring problem using Grover search and is based on the newly designed comparator. Now, MCT\_Realization algorithm takes generated circuit netlist as input  and realizes MCT gates to NISQ hardware compatible 1-qubit and 2-qubit gates \cite{portugal}. Finally SABRE \cite{sabre} algorithm has been used for mapping generated circuit by NISQ devices based on the qubit topology.

This section outlines the proposed methodology for the Oracle circuit synthesis of the $k$-Coloring problem as an application of the Grover's search algorithm. 

\subsection{Proposed Oracle for $k$-Coloring Problem}
 The quantum circuit block of Oracle for the $k$-coloring problem is shown in Figure \ref{gen}. The construction of Oracle for $k$-coloring problem is divided into five parts starting with initialization, which is essentially required in Grover's Algorithm.

\subsubsection{Initialization}
\begin{figure*}[!h]
\centering
\includegraphics[width=5in, height=0.98in]{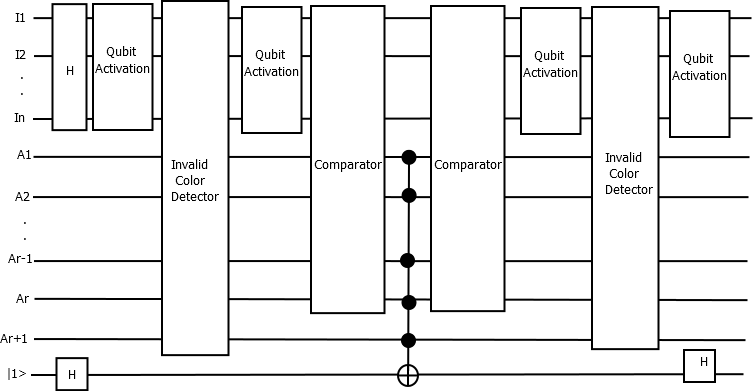}
\caption{Block Diagram of Generalized Oracular Circuit}
\label{gen}
\end{figure*}

If there are $n$ vertices, $e$ edges in the input graph and $k$ is the number of colors, then the total number of data qubits required to represent all the colored vertices are $n * \ceil{\log_2 k}$. The Oracle checks for all the right combination of properly colored vertices with $k$ or fewer colors from a combination of all possible colored vertices. Hence, a superposition of $m=n*\ceil{\log_2 k}$ qubits will generate all possible combination of colored vertices. The initial data qubits in Figure \ref{gen} include $m$ qubits prepared in the ground state $\vert\psi\rangle=\ket{0}^{\otimes m}$, due to the re-usability property of ancilla qubits, $r= n$ ancilla qubits in the exited state $\ket{\theta}=\ket{1}^{\otimes r}$ (These $r$ ancilla qubits are required to prepare Invalid Color detector block and Comparator block which are described in next subsection thoroughly), one ancilla qubit in the ground state $\ket{\zeta}=\ket{0}$ (1 ancilla is required iff invalid color exists), and one output qubit in the excited state $\ket{\phi}=\ket{1}$ is required to perform CNOT/Toffoli/MCT operation of the Oracle. This entire initialization can be mathematically  written as:

\[\vert\psi\rangle\otimes\vert\theta\rangle\otimes\vert\zeta\rangle\otimes\vert\phi\rangle=\vert 0\rangle^{\otimes m}\otimes\vert 1\rangle^{\otimes r}\otimes\vert 0\rangle\otimes\vert 1\rangle\]

\subsubsection{Hadamard Transformation}

After the initialization, the Hadamard transform $H^{\otimes m}$ on data qubits and $H$ on output qubit is performed, therefore all possible states are superposed as $\ket{\psi_{0}}\otimes\ket{\theta_{0}}\otimes\ket{\zeta_{0}}\otimes\ket{\phi_{0}}$, where

\[\vert\psi_{0}\rangle=\,{1\over\sqrt{2^{m}}}\sum_{i=0}^{2^{m}-1}\vert i\rangle\]

\[\ket{\theta_{0}}=\ket{1111 ..... r(times)}\]

\[\ket{\zeta_{0}}=\ket{0}\]

\[\vert\phi_{0}\rangle=\,{1\over\sqrt{2}}\left(\vert 0\rangle-\vert 1\rangle\right)\]

\subsubsection{Proposed $U_f$ Transformation:}

This proposed unitary $U_f$ transformation has two distinct parts.
\par \textbf{(1)Reduction of Invalid Colors:} Since $c=\ceil{\log_2 k}$, hence we consider maximum $2^c$ colors. If $2^c=k$, then all colors are valid colors, else there will be a set of $2^c -k$ invalid colors. The search space should be optimized with valid colors. This can be carried out using the following steps: 

\par \textbf{Qubit Activation:} Colors are needed to be numbered as $\{0,1,2 .... 2^{c}-1\}$. After the Hadamard transformation, the input data qubit lines act as the binary representation of combination of all possible colored vertices. But, the oracle checks only the combination of valid colors $k$. To make sure that the Oracle is checking only the $k$-colored combination of vertices, all the input qubit lines are needed to be in the excited state $\ket{1}$ for those particular combinations of invalid colors by making input qubit lines suitable as control lines for CNOT/Toffoli/MCT operation. A number of NOT gates have to be imposed on the input qubit lines, which are in the ground state $\ket{0}$ followed by the application of 'Invalid Color Detector'. This 'Qubit Activation' has to be applied again after 'Invalid Color Detector' to return back to the initial superposed quantum state.

\textbf{Invalid Color Detector:} If any invalid color is detected in any combination of colored vertices then that combination is discarded using the following function ICD (Invalid Color Detector):
\begin{equation}
\resizebox{.85\hsize}{!}{$ ICD(I_1, I_2, .., I_n, f)= \left\{\begin{array}{ll}\mbox{$f=0$,} & \mbox{if $I_1 or I_2 or.. I_n =$ Invalid color};\\
\mbox{$f=1$,} & \mbox{No invalid color}.\\
\end{array}\right. $}
\end{equation}

\begin{figure}[!h]
\centering
\includegraphics[width=40mm,height=1.6cm]{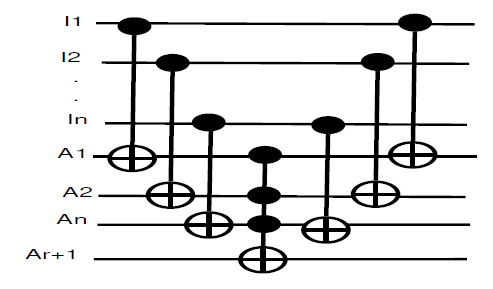}
\caption{Invalid Color Detector}
\label{ICD}
\end{figure}

Figure \ref{ICD} describes the circuit synthesis of 'Invalid Color Detector' for $n$ vertices, where $I_1, I_2, .., I_n$ are the data qubits.

\begin{figure}[!h]
\centering
\includegraphics[width=50mm,height=1.3cm]{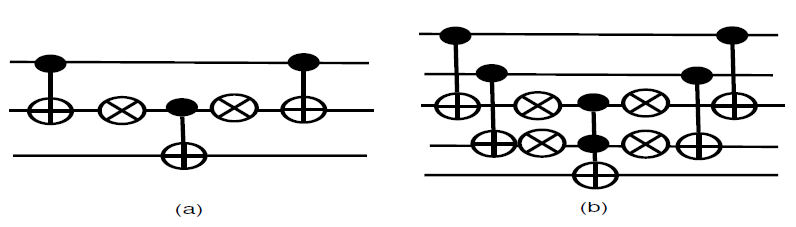}
\caption{Example Comparator: (a) 2-qubit; (b) 4-qubit}
\label{comparator}
\end{figure}

 \textbf{(2)Binary Comparator:} A newly proposed binary comparator circuit can be defined as:

\begin{equation}
\resizebox{.55\hsize}{!}{$ Comparator(a, b, f)= \left\{\begin{array}{ll}\mbox{$f=0$,} & \mbox{if $a = b$};\\
\mbox{$f=1$,} & \mbox{$a \neq b$}.\\
\end{array}\right. $}
\end{equation}

where $a$ and $b$ are the comparing inputs which represent the colored vertices of the given graph and $f$ is the ancilla
qubit. Circuit synthesis for $2$-qubit and 4-qubit  comparator is shown
in Figure \ref{comparator}. CNOT, NOT, Toffoli/MCT gates
are used to design the complete circuit synthesis for the binary comparator. 

With the help of these invalid color reduction function and newly proposed comparator, The design of $U_f$ of an Oracle for $k$-coloring problem is effectively developed.

\subsubsection{MCT Operation}
The output qubit state $\vert\phi_{0}\rangle$ is initially set as
${1\over\sqrt{2}}\left(\vert 0\rangle-\vert 1\rangle\right)$. Applying an MCT gate on the output line considering ancilla qubits as control, results in an eigenvalue kickback $-1$,
which causes a phase shift for the respective input
state/states, which helps to find out all the combination of properly colored set of vertices. The algorithm that generates the gate level synthesis of the proposed method is outlined in next subsection.

\subsection{Proposed Algorithm for Oracle Circuit Synthesis}
The proposed algorithm Algorithm 1 (AutoGenOracleK-Coloring) of automated oracular circuit synthesis for the $k$-coloring problem is illustrated in this subsection. The algorithm takes as input the adjacency matrix of the given graph and the number of colors $k$. The output of the algorithm is a circuit netlist in the form of QASM.

\begin{figure*}[!h]
\centering
\includegraphics[width=6in,height=1in]{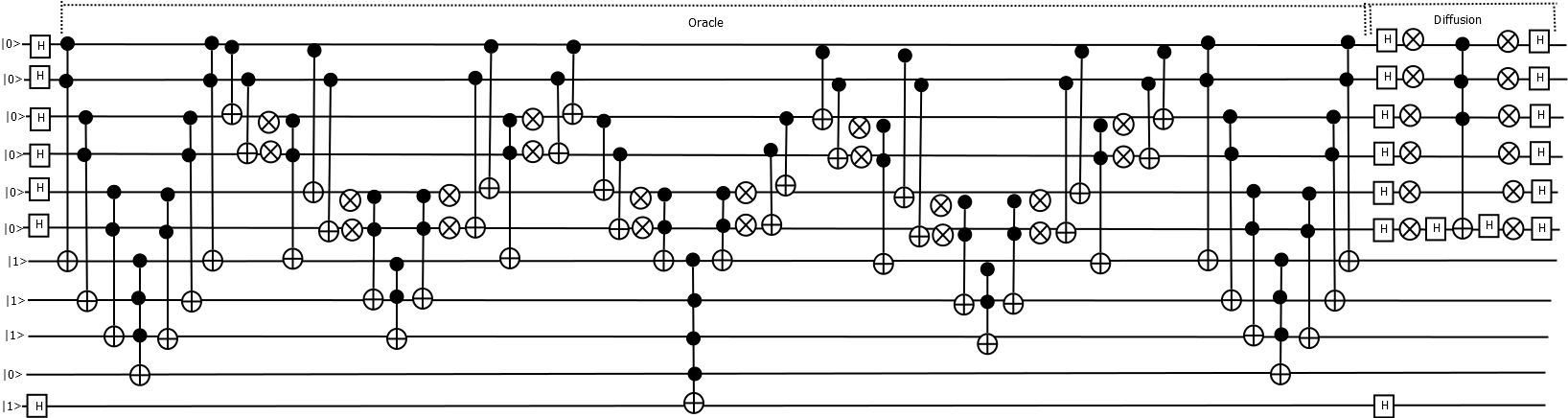}
\caption{Gate level representation of 3-coloring problem for example graph}
\label{complete}
\end{figure*}


\begin{algorithm}[!h]
\begin{algorithmic}[1]
\label{algo:one}
\caption{AutoGenOracleK-Coloring($G(V,E)$)}\label{euclid}
\STATE \bf{INPUT} : Adjacency matrix $adj(n,n)$ of graph(G) $G(V,E)$, $V=n$ and $E=e$ where, V is the set of nodes and E is the set of edges,
Number of input data qubit lines required  $I_{r}=n*\lceil \log_{2} k \rceil$(input lines for $n$ nodes and $k$ colors)$+$ ancilla lines required$= n$ $+1$ ancila line for reduction of invalid colors(if required) $+1$(output line($O$)),
$A_{r}$ represents ancilla line where, $1 \leq r \leq n$, $A_{r+1}$ represents ancilla line for invalid color (if required).
\STATE \bf{OUTPUT} : Circuit netlist (QASM)
\STATE Initialize $I_r$ input lines with $\ket{0}$ followed by Hadamard gate, ancilla lines $A_r$ with $\ket 1$, $A_{r+1}$ with $\ket 0$, and output line $O$ with $\ket 1$ followed by a Hadamard gate.

\STATE Apply Invalid Color Detector (if required) for all possible invalid colors with suitable Qubit Activation with $I_r, A_r$ as control and $A_{r+1}$ as target.

\STATE  $l \leftarrow n$, $f \leftarrow 1$
\FOR {$i\gets1$ to $n-1$}

\STATE $r \leftarrow f$, $m \leftarrow f$
 \FOR {$j\gets{i+1}$ to $n$}
\IF {$adj(i, j) \leftarrow 1$ ($i$ and $j$ are connected by an edge($e$))}
\STATE Use a comparator circuit with the input lines ($I_{i}$, $I_{j}$) corresponding $(i,j)$ as control and the ancilla line $A_{r}$  as target.
\STATE $r \leftarrow r+1$

\ENDIF
\ENDFOR
\IF {$r > f+1$}
\STATE Use a Toffoli/MCT gate with all ancilla lines $A_r$ as control and $A_l$ as target
\STATE $l \leftarrow l-1$
 \FOR {$m\gets{i+1}$ to $n$}
\IF {$adj(i, j) \leftarrow 1$ ($i$ and $j$ are connected by an edge($e$))}
\STATE Use a comparator circuit with the input lines ($I_{i}$, $I_{m}$) corresponding $(i,j)$ as control and the ancilla line $A_{m}$  as target.
\STATE $m \leftarrow m+1$

\ENDIF
\ENDFOR
\ELSIF {$r = f+1$ }
\STATE $f \leftarrow f+1$
\ENDIF
\ENDFOR
\STATE Use an MCT gate with all the ancilla lines $ A_{1}, A_{2}, \dots A_{r+1}$ as control and $O$ as output.

\STATE Repeat step 5-26.
\STATE Repeat step 4.

\end{algorithmic}
\end{algorithm}

From the details of the adjacency matrix and the number of given color, it can be easily estimated that the total number of qubit lines required to generate the Oracle circuit. All the input data qubits are initialized with $\ket 0$ followed by Hadamard, ancilla lines ($A_{r}$) are initialized with $\ket 1$, ancilla line $A_{r+1}$ is initialized with $\ket 0$ and the output line is initialized with $\ket 1$ followed by Hadamard. First of all,  apply Invalid Color Detector with suitable Qubit Activation (if invalid color exists) with $I_r, A_r$ as control and $A_{r+1}$ as the target. Then, between two adjacent vertices$(i,j)$, a comparator circuit is used with two input lines$(i,j)$ as control and the ancilla line($A_{r}$) as output and perform this same task for all the adjacent vertices. Then, an MCT gate is used with all the ancilla lines $A_{r}$ and $A_{r+1}$ as control and the output line as output for the flip operation of Grover's Oracle. To mirror everything of the Oracle circuit, we have repeated the previous steps as shown in Algorithm 1.

\subsection{Circuit Cost Estimation}

The design of generalized  Oracle for our algorithm is already described. Now, the circuit cost analysis of the oracular circuit is given in Table \ref{CC}.

\begin{table}[!h]
\centering
\caption{Circuit Cost Analysis of  Oracle}
\begin{tabular}{ |c|c|c| }
  \hline

No. of Vertex & Maximum Ancilla Required & Maximum Gate Count \\ \hline
$3$  & $3+1=4$ & $67$ \\  \hline 
$n$ & $O(n)$ & $O(n^2 *log_{2}n)$ \\ \hline
\end{tabular}
\label{CC}
\end{table}

  For $n$-vertices graph and $k$ given color, $n* \lceil log_{2} k \rceil$ data qubits are required. For $n$-vertices graph, at most $n+1$ number of  ancilla are needed and at most $O(n^2 * log_{2} n)$ gates are required to design the oracular circuit. The gate-optimized circuit synthesis of the 3-coloring problem for example graph of three vertices with three connected edges ($K_{3}$) is shown in Figure \ref{complete}.

\subsection{Diffusion}
The second part of Grover's algorithm is the circuit implementing the function of diffusion. When the operation is applied to a superposition state, it actually keeps the component in the $\ket{\psi_{0}}$ direction unchanged, while inverting the components in dimensions that are perpendicular to $\ket{\psi_{0}}$. This can be represented as 

\[I_{\left\vert\psi_{0}^{\perp}\right\rangle}=-I_{\vert\psi_{0}\rangle}\]

where, \[\vert\psi_{0}\rangle=\,{1\over\sqrt{2^{n}}}\sum_{i=0}^{2^{n}-1}\vert i\rangle\]

The diffusion operator is a unitary matrix. The general matrix for the diffusion operator for an $d$-dimensional quantum system is shown below:

\begin{equation*}
diff_d = \left( \begin{matrix}
    \frac{2}{d}-1 & \frac{2}{d} & \frac{2}{d} & \ldots & \frac{2}{d} \\
    \frac{2}{d} & \frac{2}{d}-1 & \frac{2}{d} & \ldots & \frac{2}{d} \\
    \frac{2}{d} & \frac{2}{d} & \frac{2}{d}-1 & \ldots & \frac{2}{d} \\
    \vdots & \vdots & \vdots & \ddots & \vdots \\
    \frac{2}{d} & \frac{2}{d} & \frac{2}{d} & \ldots &  \frac{2}{d}-1 \\
\end{matrix} \right)
  \end{equation*}

\begin{figure*}[!h]
\centering
\includegraphics[width=6.5in, height=1.7in]{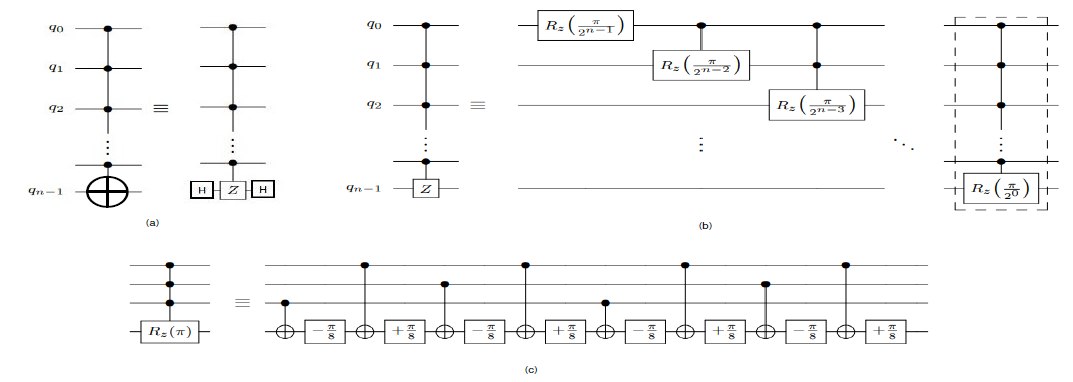}
\caption{(a) Decomposition of MCT to MCZ; (b) Decomposition of MCZ to MC$R_{x}(\pi)$ (c) Decomposition of 4-control $R_{x}(\pi)$ gate}
\label{mctreal}
\end{figure*}

\section{Mapping of $k$-Coloring Problem to NISQ Devices} 
This section focuses on the mapping of generated Oracle circuit to NISQ devices through MCT realization and SABRE algorithm for qubit mapping.
 
\subsection{Realization of MCT Gate}

Figure \ref{mctreal} shows how to decompose MCT gate to NISQ compatible 1-qubit and 2-qubit gates \cite{portugal}. Firstly MCT gate needs to be decomposed to MCZ gate. Then, the realization of MCZ gate into MC$R_{x}(\pi)$ is performed. Lastly, MC$R_{x}(\pi)$ is reduced to 1-qubit and 2-qubit gates without using any ancilla qubit.

\subsection{Qubit Mapping to NISQ Devices}

 Since, our proposed quantum circuit is logical, hence there is no constraint of qubit connectivity. For NISQ devices, there exists a specific qubit topology or coupling graph. Coupling graph defines the interaction between two physical qubits. This varies for different NISQ devices. Thus, it is obvious that mapping the logical circuit to the physical one is a challenge. The solution to this problem is the insertion of SWAP gates between the two qubits to satisfy the hardware constraint without compromising on the logic of the quantum circuit. The idea of a good qubit mapping problem is to minimize the number of SWAP insertion gates and minimize the depth of the circuit. Li et. al. proposed SWAP-based BidiREctional heuristic search algorithm (SABRE) in \cite{sabre}, which is a benchmark, since it deals with any arbitrary qubit topology for any NISQ device. Mainly three features make SABRE stand out. Firstly, it doesn’t perform an exhaustive search on the entire circuit, but it performs a SWAP-based heuristic search considering the qubit dependency. It then optimizes the initial mapping using a novel reverse traversal technique. Last but not the least, the introduction of the decay effect for enabling the trade-off between the depth and the number of gates of the entire algorithm. We use SABRE protocol so that our proposed circuit can easily be mapped to any arbitrary qubit topology.

\subsection{Experimental result of $k$-coloring Problem in NISQ Device}

 As shown in Figure \ref{complete}, the generated oracle circuit for example graph has been taken as an example case for the simulation of $k$-coloring problem which is performed on IBMQ cloud based physical device \cite{4}.
 

 \begin{figure}[!h]
\centering
\includegraphics[width=85mm, height=1.5in]{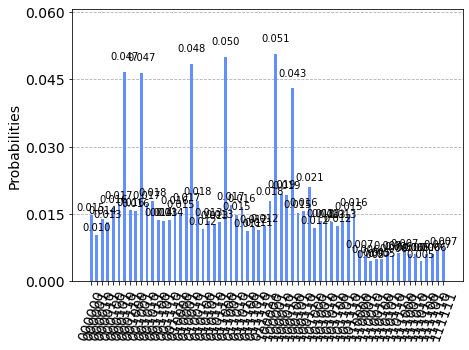}
\caption{Amplitudes of Quantum States}
\label{output}
\end{figure}

 The resultant output after applying Grover's operator is shown in Figure \ref{output}, where the amplitude of the solution state has been amplified. The location of the solution states are $\ket{011000}$, $\ket{100100}$, $\ket{000110}$,  $\ket{010010}$, $\ket{001001}$, and $\ket{100001}$ where $00$, $01$, and $10$ are the valid colors as we take $11$ as invalid color. These are the properly colored vertex combinations in the given example graph that solves the $k$-coloring problem with high probability.

\subsection{Comparative Analysis}

As compared to \cite{ibm_graphcolor}, our proposed comparator-based oracle gives better result with respect to data qubit and ancilla qubit as $n * \lceil \log_{2} k \rceil$ and $ O(n)$ respectively. Table \ref{compare} shows the comparative analysis.

\vspace{-.2cm}

\begin{table}[!htb]
\centering
\caption{Comparative Analysis}
\begin{tabular}{ |c|c|c|}
  \hline

Parameters & Hu et. al. \cite{ibm_graphcolor} & This work \\ \hline
Data Qubit Cost & $n*k$ & $n * \lceil \log_{2} k \rceil$\\
 \hline
 Ancilla Qubit Cost & $O((n*k)^2)$ & $ O(n)$\\
 \hline
 Processor & IBMQ & Any NISQ Device\\ \hline
\end{tabular}
\label{compare}
\end{table} 

\section{Conclusion}
In this paper, we have proposed an automated end-to-end framework which includes mapping of $k$-coloring problem to any NISQ devices through automatic generation of Oracle circuit using Grover search taking into account any undirected and unweighted given graph and the number of given colors ($k$), automatic MCT realization and automatic qubit mapping using SABRE for given qubit topology. Our comparator-based approach has outperformed the reduction-based approach from 3-SAT problem to 3-Color problem quite convincingly. The data qubit cost has been reduced to $n * \lceil log_2 {k} \rceil$ whereas it was $n * k$. This leads to a reduction of query complexity from $O(n * k)$ to $O(n * log_2 {k})$. In future, the swap-operation can be used for re-usability of the qubits for further optimization while generating the Oracle circuit.




\section*{Acknowledgment}

This work has been supported by the grant from CSIR, Govt. of India, Grant No. 09/028(0987)/2016-EMR-I.



%


\end{document}